\documentclass[prd,eqsecnum,preprintnumbers,twocolumn,showpacs]{revtex4}
\usepackage{amsmath,bm,graphicx,slashed}

\newcommand{\nn}{\nonumber} 
\newcommand{\beq}{\begin{equation}}
\newcommand{\eeq}{\end{equation}} 
\newcommand{\beqa}{\begin{eqnarray}} 
\newcommand{\eeqa}{\end{eqnarray}}

\def\be{\begin{equation}}
 \def\ee{\end{equation}}
 \def\bea{\begin{eqnarray}}
 \def\eea{\end{eqnarray}}
 \def\bean{\begin{eqnarray*}}
 \def\eean{\end{eqnarray*}}
\newcommand{\ie}{{\it i.e.}}

\newcommand{\morder}[1]{{\cal O}\left(#1 \right)}
\newcommand{\eq}[1]{(\ref{#1})}

\newcommand{\kvec}{{\bf k}}

 \newcommand{\gsim}{\gtrsim}

\def\COMMENT#1{}

 \def\l{\left}
 \def\r{\right}

 \def\esim{\,\mathrel{\rlap{\lower0.2em\hbox{$-$}}\raise0.15em\hbox{\footnotesize $\hskip0.04em\sim$}}\,}
 \def\gsim{\mathrel{\rlap{\lower0.2em\hbox{$\sim$}}\raise0.2em\hbox{$>$}}}
 \def\ksim{\mathrel{\rlap{\lower0.2em\hbox{$\sim$}}\raise0.2em\hbox{$<$}}}

\begin{document}

%==========================================================
%\preprint{}
\title{Collisional energy loss of a fast heavy quark in a quark-gluon plasma} 

\author{St\'ephane Peign\'e}
%% \email{peigne@subatech.in2p3.fr}

\author{Andr\'e Peshier}
%% \email{peshier@subatech.in2p3.fr}

\affiliation{SUBATECH, UMR 6457, Universit\'e de Nantes \\ Ecole des
Mines de Nantes, IN2P3/CNRS. \\ 4 rue Alfred Kastler, 44307 Nantes cedex 3, France}

\date{\today}

\begin{abstract}
We discuss the average collisional energy loss $dE/dx$ of a heavy quark crossing a quark-gluon plasma, in the limit
of high quark energy $E \gg M^2/T$, where $M$ is the quark mass and $T \ll M$ is the plasma temperature.
In the fixed coupling approximation, at leading order $dE/dx \propto \alpha_s^2$, 
with a coefficient which is logarithmically enhanced. The soft logarithm arising from 
$t$-channel scattering off thermal partons is well-known, but a collinear logarithm 
from $u$-channel exchange had previously been overlooked. We also 
determine the constant beyond those leading logarithms. 
We then generalize our calculation of $dE/dx$ to the case of running coupling. 
We estimate the remaining theoretical uncertainty of $dE/dx$, which turns out to be quite large under RHIC conditions.
Finally, we point out an approximate relation between $dE/dx$ and the QCD Debye mass, from which we derive an upper bound to $dE/dx$ for all quark energies. 
\end{abstract}
\maketitle

%==========================================================
\section{Introduction}

Jet quenching, as first anticipated by Bjorken \cite{bj}, is a crucial
probe of the state of matter created at the Relativistic Heavy Ion Collider (RHIC). 
The quenching of light hadron spectra at large $p_{\perp}$ \cite{phenix,star} is usually attributed 
to the radiative energy loss of the (light) parent parton when crossing the hot or dense medium.
On the other hand, recent data on heavy flavour quenching 
\cite{Adler:2005xv,Bielcik:2005wu} suggest that for {\it heavy} quarks, purely radiative energy loss 
{\it might} be insufficient to explain the observed attenuation. 
This has renewed the interest in the {\it collisional} part $\Delta E_{coll}$ 
of the parton energy loss, in particular in the case of a heavy quark \cite{Wicks:2005gt}. 
It is not clear yet whether such a collisional contribution can help understanding the data
on heavy flavour quenching. For instance, although some studies support the possibility of
a quite large collisional loss \cite{Mustafa:2003vh,Mustafa:2004dr,DuttMazumder:2004xk},
a recent comparison between collisional and radiative losses \cite{Zakharov:2007pj} indicates that 
the collisional contribution to parton energy loss might be small ($\sim 20\%$) compared to the radiative 
one, even for heavy quarks. It should however be noted that a relatively small 
{\it average} collisional loss might be compatible with an important effect of 
collisions on quenching \cite{Qin:2007rn,Peshier:2008bg}, due to differently behaved collisional and radiative 
energy loss {\it probability distributions}.

In any case, in the present (unclear) situation where the importance 
of collisional energy loss is reconsidered, we believe it is not useless to state 
the correct result for the (average) collisional loss of an energetic heavy quark. 

A basic quantity required to study collisional quenching is the rate of energy loss per unit distance, 
$dE/dx$, of a parton produced in the remote past and travelling in a large size medium, 
as studied in Refs.~\cite{bj,TG,BTqcd}. 
For heavy ion collisions, where a parton initially produced in a hard subprocess crosses a medium of finite size $L$, 
we expect deviations from the linear law $\Delta E_{coll}(L) = (dE/dx) \cdot L$ \cite{Gossiaux:2007gd}. 
Nonetheless, knowing the `asymptotic' rate $dE/dx$ is a prerequisite before attempting a more refined evaluation of 
$\Delta E_{coll}$, including in particular finite size effects. 

So far, the most detailed calculation of $dE/dx$ for a heavy quark in a quark-gluon plasma (QGP) is due to Braaten and 
Thoma (BT) \cite{BTqcd}, and is based on their previous QED calculation of $dE/dx$ for a muon \cite{BTqed}.
As we recently analyzed \cite{PP}, the BT QED calculation relies on an assumption
on the momentum exchange $q$ in elastic scattering, namely $q \ll E$, which is incorrect in the domain $E \gg M^2/T$. 
Therefore, the BT results for $dE/dx$, both in QED \cite{BTqed} and in QCD \cite{BTqcd}, need to be corrected 
in this limit. This was done in \cite{PP} in the QED case. Here we consider the QCD case of a fast heavy quark.

We assume the heavy quark to be produced in the hard partonic subprocess of the heavy-ion collision, \ie\ 
to be present in the 'initial' stage of the QGP evolution. It then crosses the QGP (supposed thermally equilibrated) 
by losing some energy, before hadronizing (within a jet) into a heavy-flavoured hadron.
The situation we consider should be appropriate when dealing with heavy-quark tagged jets, 
where the energy loss is defined as the difference between the initial and final energies
of the {\it flagged} heavy quark. This is different from light parton (or untagged 
heavy quark) energy loss, which should be defined (at the partonic level) as the energy difference 
between the {\it leading} 
partons in the initial and final states. Indeed, without tagging, the flavour or even the 
type (quark or gluon) of the initial energetic parton does not have to be conserved in the `energy loss' process. 

In section \ref{sec:fixedcoupling} we study the fixed coupling approximation 
which allows us to closely follow the lines of our QED calculation
\cite{PP}. In section \ref{sec:runningcoupling}
we show how the result with a running coupling can be simply inferred, and discuss the theoretical
uncertainty. The latter is illustrated by a numerical estimate in section \ref{sec:numest}.

%==========================================================
\section{Fixed coupling approximation}
\label{sec:fixedcoupling}

In this section we assume a fixed coupling, which is justified when the running coupling does not change much in 
the range of probed momenta (we will specify this condition below). 
We calculate, in the limit $E \gg M^2/T$, the heavy quark collisional loss $dE/dx$ beyond logarithmic accuracy. 
For the sake of clarity we first focus on the leading logarithms before calculating the constant next to those logarithms.

\subsection{Leading logarithmic terms}
\label{IIA}

The leading logarithms in $dE/dx$ for a heavy quark in the quark-gluon plasma have the same origin as in the QED analog of a 
muon crossing an $e^\pm \gamma$ plasma. 
We thus make a detour to QED and outline the derivation of \cite{PP}, which is straightforward at leading logarithmic accuracy.
The QCD result for $dE/dx$ of a heavy quark will then be obtained from simple considerations. 

\subsubsection{Muon energy loss in a QED plasma}
At leading order the energy loss of a muon (of mass $M$ and energy $E$) arises from elastic scattering off 
thermal electrons or positrons (Fig.~\ref{fig:amp}a) and photons (Figs.~\ref{fig:amp}b-c). Scattering 
off electrons corresponds to $t$-channel exchange, whereas scattering off photons (Compton scattering) receives
contributions from $s$ and $u$-channels. The muon energy loss is given by \cite{BTqed}

\begin{widetext} %-----------------------------
\begin{figure*}[t]
\centering
\begin{tabular}{c@{\hspace*{12mm}}c@{\hspace*{8mm}}c@{\hspace*{8mm}}c}
    \includegraphics[scale=0.7]{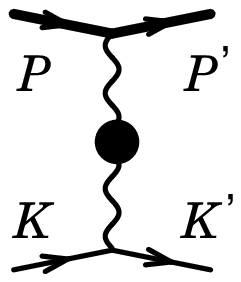}
   &
   \includegraphics[scale=0.7]{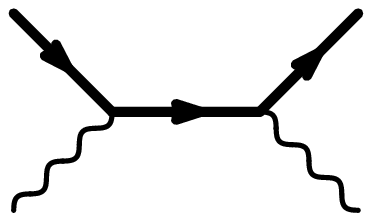}
   &
    \includegraphics[scale=0.7]{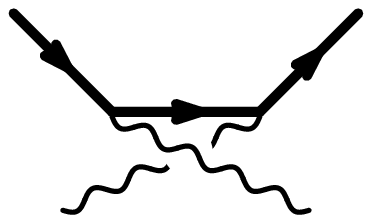}
   &
    \includegraphics[scale=0.7]{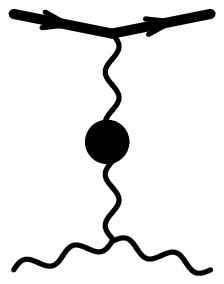}
   \\
     (a) & (b) & (c) & (d)
   \end{tabular}
   \caption{Amplitudes for heavy muon (quark) elastic scattering in a QED (QCD) plasma. A curly line
denotes a photon (QED) or a gluon (QCD). The amplitude (d) is specific to the QCD case.
The blob in (a) and (d) denotes the resummed hard thermal loop boson propagator, which is
necessary to screen the $t$-channel contribution in the infrared.}
  \label{fig:amp}
  \end{figure*} 

\begin{align}
  \frac{dE^{(\mu)}}{dx}
  = \sum_i
  \frac{1}{2Ev}
	\int_k \frac{n_i(k)}{2k}
	\int_{k'} \frac{\bar n_i(k')}{2k'}\,
	\int_{p'} \frac1{2E'}\,	(2\pi)^4\delta^{(4)}(P+K-P'-K')\,
	\frac{1}{d}\sum_{\rm spins}\l| {\cal M}_i\r|^2\, \omega  \, ,
\label{eq:dEdx0}
\end{align}
where $\omega = E-E'$. 
The tree-level amplitude ${\cal M}_i$ corresponds to scattering off a thermal particle of type
$i = e^+, e^-, \gamma$. Each $|{\cal M}_i|^2$ is summed over initial and final spin states, 
and we divide by the degeneracy factor $d = 2$ of the incoming muon. 
Furthermore, $n_i(k) = (\exp(k/T) \mp 1)^{-1}$ is the thermal distribution of the target particles, 
and $\bar n_i = 1 \pm n_i$ accounts for the Bose enhancement or Pauli blocking for the scattered state. 
We also use the shorthand notation $\int_k \equiv \int d^3 \kvec /(2\pi)^3$. 
\end{widetext} % --------------------

In the $E \gg T$ limit, \eq{eq:dEdx0} can be simplified to \cite{PP}
\beq 
\frac{dE^{(\mu)}}{dx}  =  \sum_i d_i \int_k \frac{n_i(k)}{2k}
	\int_{t_{\rm min}}^{t_{\rm max}} dt\, (-t)\, \frac{d\sigma_i}{dt} \, ,
\label{eq:dEdx_asy}
\eeq
where $d_i$ is the degeneracy factor of the target particles, and 
\beq
	\frac{d\sigma_i}{dt}
	=
	\frac1{16\pi {\tilde s}^2}\,
	\frac1{d\, d_i} \sum_{\rm spins}\l| {\cal M}_i \r|^2 
\label{crossdef}
\eeq
is the corresponding differential cross section.
We also define $\tilde s \equiv s -M^2$, as well as for later reference 
$\tilde u \equiv u -M^2$. In Eq.~\eq{eq:dEdx_asy} the bounds on $t$ are set by kinematics, 
$t_{\rm min} = -{\tilde s}^2/s$, and $t_{\rm max} = 0$. 
We will focus on the limit $E \gg M^2/T$, which implies
$s = (P+K)^2 = M^2 + 2 PK \sim \morder{E T} \gg M^2$, so that $t_{\rm min} \simeq - s$. 

To obtain the leading logarithm from the $t$-channel contribution we can assume $|t| \ll s$ and approximate
\beq 
\sum_{i= e^\pm} \sum_{\rm spins} \l| {\cal M}_{i} \r|^2 \simeq 32\,e^4 \frac{{\tilde s}^2}{t^2} .
\label{tampsquared}
\eeq
This contributes to \eq{eq:dEdx_asy} as
\beq
\frac{e^4}{\pi} \int_k \frac{n_F(k)}{2k} \int_{t_{\rm min}}^{t_{\rm max}} \frac{dt}{-t} \simeq 
\frac{e^4 T^2}{48 \pi} 
\left[ \ln\frac{E T}{m_D^2} + \morder{1} \right] .
\label{qed-t}
\eeq
To obtain the r.h.s., we replaced $t_{\rm min} \simeq -s \to -ET$, which is justified to logarithmic accuracy. 
Strictly speaking, with $t_{\rm max} = 0$ the integral in \eq{qed-t} would be infrared divergent.
As is well-known, this divergence is screened by medium effects. To logarithmic accuracy 
it is sufficient to take $t_{\rm max} = - m_D^2$ as an effective infrared cut-off, 
where $m_D = eT/\sqrt{3}$ is the Debye screening mass in the QED plasma \cite{BI}. The logarithm 
in \eq{qed-t} thus arises from the domain $m_D^2 \ll |t| \ll s \sim ET$. 
A more accurate description of screening requires using the resummed hard thermal loop (HTL) \cite{HTL,BI} 
photon propagator in the $t$-channel (instead of the bare one with effective cut-off), as pictured in Fig.~\ref{fig:amp}a. 
This is needed to control the constant $\sim \morder{1}$ 
in \eq{qed-t}, as will be recalled in section \ref{IIB}. Finally, in order to obtain \eq{qed-t} we used 
\beq
\int_k \frac{n_F(k)}{2k} = \frac{T^2}{48}  \, .
\label{intkfermion}   
\eeq

The contribution from Compton scattering to the muon energy loss brings another logarithm, 
arising from the square of the $u$-channel amplitude, more specifically from the domain 
$\tilde u_{\rm min} \ll \tilde u \ll \tilde u_{\rm max}$ 
\cite{PP}. In this domain the Compton scattering amplitude squared can be approximated as 
\beq
\sum_{\rm spins}\l| {\cal M}_\gamma \r|^2 \simeq 8e^4 \frac{\tilde s}{-\tilde u} \, .
\label{eq:M2Comptonappr}
\eeq
Changing variables from $t$ to $\tilde u$ in \eq{eq:dEdx_asy}, and using the bounds
$\tilde u_{\rm min} = -\tilde s \simeq - s$ and $\tilde u_{\rm max} = -M^2\, \tilde s/s \simeq -M^2$, 
\eq{eq:M2Comptonappr} contributes to $dE/dx$ as 
\beq
\frac{e^4}{4\pi} \int_k \frac{n_B(k)}{2k} \int_{\tilde u_{\rm min}}^{\tilde u_{\rm max}} 
\frac{d\tilde u}{-\tilde u} \simeq 
\frac{e^4 T^2}{96 \pi} \left[ \ln\frac{E T}{M^2} + \morder{1} \right] \, , 
\label{qed-u} 
\eeq
where we used 
\beq
\int_k \frac{n_B(k)}{2k} = \frac{T^2}{24}  \, . 
\label{intkboson} 
\eeq
Adding \eq{qed-t} and \eq{qed-u} we obtain the muon energy loss to logarithmic accuracy,
\beq
\frac{dE^{(\mu)}}{dx} = \frac{e^4 T^2}{48 \pi} \left[ \ln\frac{E T}{m_D^2} + \frac{1}{2}\ln\frac{E T}{M^2} + \morder{1} \right] \, .
\label{qed-tot}
\eeq

\subsubsection{Heavy quark energy loss in a QGP}

The heavy quark collisional energy loss arises from elastic scattering off 
thermal quarks (Fig.~\ref{fig:amp}a) and gluons (Figs.~\ref{fig:amp}b-d). Compared to the 
QED case, there is one additional amplitude (Fig.~\ref{fig:amp}d), corresponding 
to $t$-channel exchange off thermal gluons. Using the cross sections \cite{Combridge:1978kx} corresponding to the 
QCD amplitudes of Fig.~\ref{fig:amp} (with tree level gluon propagators in the $t$-channel amplitudes), 
we can easily identify the origin of a logarithmic enhancement 
in the QCD analog of \eq{eq:dEdx_asy}. There is a soft logarithm $\sim \int dt/t$ 
arising from $|{\cal M}_{\rm q}|^2$ and $|{\cal M}_{\rm g}^t|^2$, and a 
collinear logarithm $\sim \int d{\tilde u}/{\tilde u}$ from $|{\cal M}_{\rm g}^u|^2$.
The interference terms are not enhanced by any logarithm -- they however contribute to the constant 
to be evaluated in section \ref{IIB}. 

The heavy quark energy loss is readily derived from the QED result as follows. The contribution
from $|{\cal M}_{\rm q}|^2$ is obtained by multiplying \eq{qed-t} 
by the number of quark flavours $n_f$ 
and by the color factor $(N_c^2-1)/(4N_c) = 2/3$. 
To get the $t$-channel contribution from $|{\cal M}_{\rm g}^t|^2$ for
scattering off thermal gluons, we also start from \eq{qed-t} and multiply by $1/2$ since contrary to 
the electron, a gluon is its own antiparticle. This factor is compensated by a factor $2$ arising from 
the difference between bosons and fermions when performing the 
integral over $k$ (compare \eq{intkfermion} and \eq{intkboson}). The color 
factor for this contribution is $(N_c^2-1)/2 = 4$. 
Finally, the $u$-channel contribution from $|{\cal M}_{\rm g}^u|^2$ 
is obtained from \eq{qed-u} by multiplying by the color factor $C_F^2 = 16/9$. 
Introducing the QCD coupling by $e^2 \to g^2 = 4 \pi \alpha_s$, we obtain the 
heavy quark energy loss in a QGP by summing all contributions,
\beq
\frac{dE}{dx}  = \frac{4 \pi \alpha_s^2 T^2}{3} 
\left[ \left( 1 +\frac{n_f}{6} \right) \ln\frac{E T}{m_D^2} + \frac{2}{9} \ln\frac{E T}{M^2} 
+ c(n_f) \right] \, .
\label{qcd-tu}
\eeq
Here $m_D^2 = 4 \pi \alpha_s T^2 (1 + n_f/6)$ \cite{BI} is the Debye mass squared in the QGP. 
The constant $c(n_f) \sim \morder{1}$ is evaluated in the next section, see \eq{cofnf}.

\subsection{The constant beyond leading logarithms}
\label{IIB}

In the QED case, the constant $\sim \morder{1}$ in \eq{qed-tot} was determined in Ref.~\cite{PP}. 
Here we infer from the QED calculation the constant $c(n_f)$ appearing in the QCD expression \eq{qcd-tu}.

The logarithms $\ln(E T/m_D^2)$ and $\ln(E T/M^2)$ arise from the ranges
$m_D^2 \ll |t| \ll s$ and $M^2 \ll |\tilde u| \ll s$, with $s \sim ET$. This is why they could be easily 
obtained in section \ref{IIA}, using approximate expressions for the squared amplitudes (see \eq{tampsquared}
and \eq{eq:M2Comptonappr}) in those kinematical domains. Controlling the constant next to the leading logarithms, 
however, requires considering the complete phase space $0 \leq |t| \leq |t_{\rm min}| \simeq s$
and $|\tilde u_{\rm max}| \simeq M^2 \leq  |\tilde u| \leq |\tilde u_{\rm min}| \simeq s$.

In order to treat correctly the domain $|t| \sim m_D^2$, it is convenient to introduce 
an intermediate scale $t^\star$ chosen as $m_D^2 \ll |t^\star| \ll T^2$ \cite{PP}.
The contribution to $dE/dx$ from $|t| < |t^\star|$ is determined by the HTL self-energy of the exchanged gluon 
in Figs.~\ref{fig:amp}a and \ref{fig:amp}d, similarly to 
the QED case where it depends on the photon HTL self-energy (and where only Fig.~\ref{fig:amp}a contributes). 
In both cases, the HTL self-energies have the same form, up to the 
replacement of the QED Debye mass by its QCD counterpart. Introducing the overall color factor $C_F =4/3$, 
the QCD result is thus inferred from \cite{PP} to be
\beqa
\left. \frac{dE}{dx} \right|_{|t| < |t^\star|} &=& \frac{\alpha_s m_D^2}{3} \left[ \ln\frac{|t^\star|}{m_D^2} + \ln{2} \right]\, .
\label{tsoft}
\eeqa
We stress that the latter equation is valid {\it beyond} logarithmic accuracy, \ie\ the constant 
next to the leading logarithm (written as $\ln{2}$ here) is meaningful. 

The contribution from $|t| > |t^\star|$ can be evaluated by using tree level internal propagators
in the amplitudes of  Fig.~\ref{fig:amp}. (Since  $|\tilde u| \geq M^2 \gg T^2$, HTL corrections 
to the internal quark propagator in Figs.~\ref{fig:amp}b and \ref{fig:amp}c are irrelevant.) 
However, in order to control the constant beyond the leading logarithms, accurate 
expressions for the tree level cross sections have to be used. 

For the contribution from scattering off light quarks (Fig.~\ref{fig:amp}a), the calculation is similar
to the QED case, which is done by using 
\beq 
\sum \l| {\cal M}_q \r|^2 \propto \left[ \frac{{\tilde s}^2}{t^2} + \frac{s}{t} +\frac{1}{2} \right] 
\label{exactQqampsquared}
\eeq
instead of \eq{tampsquared}, and by performing the integrals in \eq{eq:dEdx_asy} with 
$|t^\star| \leq |t| \leq s$. The QCD result reads:
\beqa
\left. \frac{dE_q}{dx} \right|_{|t| > |t^\star|} 
= \frac{4 \pi \alpha_s^2 T^2}{3} \frac{n_f}{6} \left[ \ln\frac{8ET}{|t^\star|} -\frac{3}{4} +c \right] ,
\label{qthard}
\eeqa
where $c = \zeta'(2)/\zeta(2) - \gamma \simeq -1.147$, with $\gamma \simeq 0.577$ being Euler's constant. 

Similarly, scattering off gluons (Figs.~\ref{fig:amp}b-d) should in principle 
be evaluated with the exact tree level cross section \cite{Combridge:1978kx}. 
However, in the limit $s \sim ET \gg M^2$ we are considering, this cross section 
can be approximated as
\beq 
\sum \l| {\cal M}_g \r|^2 \propto \left[ \frac{{\tilde s}^2}{t^2} + \frac{s}{t} +\frac{1}{2}  \right]
+ \frac{2}{9}\left[ \frac{-{\tilde u}}{{\tilde s}} + \frac{{\tilde s}}{-{\tilde u}} \right] \, .
\label{Qgampsquared}
\eeq
The contribution of the first term of \eq{Qgampsquared} to 
$dE/dx$ is evaluated as for scattering off quarks (see \eq{exactQqampsquared}), except for the
factor $n_B(k)$ instead of $n_F(k)$ in \eq{eq:dEdx_asy}. 
Up to the color factor $2/9$, the second term of \eq{Qgampsquared} has the same form as 
QED Compton scattering, and its contribution to $dE/dx$ can be directly obtained from \cite{PP}.
Summing the two contributions we get
\beqa
\left. \frac{dE_{g}}{dx} \right|_{|t| > |t^\star|} 
&=& \frac{4 \pi \alpha_s^2 T^2}{3} \left[ \left( \ln\frac{4ET}{|t^\star|} -\frac{3}{4} +c \right) \right. \nn \\
&& \hskip .2cm + \left. \frac{2}{9} \left( \ln\frac{4ET}{M^2} -\frac{5}{6} +c \right) \right] \, .
\label{ghard}
\eeqa

Adding now the contributions from \eq{tsoft}, \eq{qthard} and \eq{ghard} we obtain for the constant 
$c(n_f)$ in \eq{qcd-tu}
\beq
c(n_f) = a\, n_f +b \simeq 0.146 \, n_f + 0.050   \, ,
\label{cofnf}
\eeq
where the exact values of $a$ and $b$ are 
$a = (2/3) \ln{2} -1/8 +c/6$ and $b= (31/9) \ln{2} -101/108 +11c/9$.

%==========================================================
\section{Implementing running coupling}
\label{sec:runningcoupling}

Implementing the running of $\alpha_s$ in the calculation of $dE/dx$ has already been done in Ref.~\cite{APrunning} for 
the logarithmically enhanced contribution from $t$-channel exchange. Here we will generalize the argument to the collinear 
logarithm stemming from the $u$-channel. 
We also discuss the influence of running on the term  
$\propto \alpha_s^2 c(n_f)$ in \eq{qcd-tu}, which allows us to define precisely 
the level of accuracy of our final result, see \eq{qcd-running}.

We first note that the QCD result \eq{qcd-tu} (together with \eq{cofnf}) obtained with a fixed coupling suffers from 
a lack of predictability. Indeed, the fixed coupling calculation does not specify 
at which scale $\alpha_s$ should be evaluated. As in PQCD calculations at zero temperature, 
`fixing the scale' in $\alpha_s$ demands to calculate the next order in the perturbative series
for the observable of interest. This next-to-leading order (NLO) generally brings large logarithms,
which can be `absorbed', via renormalization, by setting the scale of $\alpha_s$ in the leading order 
result. In the case of $t$-channel scattering (Figs.~\ref{fig:amp}a and \ref{fig:amp}d), the {\it vacuum} contributions to 
the self-energy and vertex corrections conspire to yield a logarithmic dependence on the invariant transfer $t$. 
In order to avoid a large NLO contribution, 
one must set the scale of $\alpha_s$ to $\sim \morder{|t|}$ 
\footnote{Setting this scale in the factor $\alpha_s^2$ of the $2 \to 2$ cross section amounts to perform a specific 
all-order resummation, corresponding physically to the final heavy quark being accompanied by collinear gluon radiation. 
The true {\it exclusive} $2 \to 2$ cross section involving charged (coloured) particles vanishes \cite{Dokshitzer:1991wu}.}. 

In order to obtain a predictive result for the $t$-channel contribution to $dE/dx$, 
we consider the term derived in section \ref{IIA}, 
which is logarithmically enhanced with a fixed coupling, and recalculate it with a 
running coupling $\alpha_s(|t|)$. The procedure is rather trivial,
\beq
 \alpha_s^2 \int^{ET}_{m_D^2} \frac{d|t|}{|t|} 
\longrightarrow 
\int^{ET}_{m_D^2} \frac{d|t|}{|t|} \alpha_s^2(|t|) \, .
\label{rule1}
\eeq 
Using 
\beq
\alpha_s(|t|) = \left[ 4\pi \beta_0 \ln{(|t|/\Lambda^2)}\right]^{-1} \, ,
\label{running}
\eeq
where $\beta_0 = (11-\frac{2}{3} n_f)/(4\pi)^2$ is the leading coefficient of the QCD $\beta$-function, 
we see that using a running coupling amounts to perform the replacement 
\beq
\alpha_s^2 \ln{\frac{ET}{m_D^2}}
\,\longrightarrow\,  
%% \frac{\alpha_s(|t_{\rm max}|)- \alpha_s(|t_{\rm min}|)}{4\pi \beta_0} 
\alpha_s(m_D^2)\alpha_s(ET) \ln{\frac{ET}{m_D^2}}  \, .
\label{rule2}
\eeq
As noted in \cite{APrunning}, the latter result becomes $E$-independent in the $E \to \infty$ limit, 
where the logarithmic enhancement of the fixed coupling result \eq{qcd-tu} is invalid. 

The accuracy of 
the above procedure is inferred by noting that the scale of $\alpha_s$ in the r.h.s.\
of \eq{rule1} can be in principle chosen as $C |t|$ rather than $|t|$, with $C$ a constant of order unity. 
This implies an ambiguity 
of order $\alpha_s^3 \ln{ET/m_D^2}$ to the r.h.s.\ of 
\eq{rule2} \footnote{Using the two-loop running coupling instead of the one-loop expression 
\eq{running} yields similar corrections, provided we assume $\ln{\ln{(|t|/\Lambda^2)}} \sim \morder{1}$.}.
Thus specifying the term $\propto \alpha_s^2 c(n_f)$ in \eq{qcd-tu} is meaningful provided we assume
\beq
\alpha_s(m_D^2) \ln{\frac{ET}{m_D^2}} \ll 1 \;\Longleftrightarrow\; \ln{\frac{ET}{m_D^2}} 
\ll \ln{\frac{m_D^2}{\Lambda^2}}  \, .
\label{approx}
\eeq

For the $u$-channel contribution (Fig.~\ref{fig:amp}c) to the cross section, $\alpha_s$ should be evaluated at a scale
$\sim \morder{|{\tilde u}|}$. Similarly to the above discussion, the logarithmic integral appearing in 
\eq{qed-u} is modified to $\sim \int d|\tilde u| \alpha_s^2(|\tilde u|)/|\tilde u|$, and the 
collinear logarithm in \eq{qcd-tu} becomes
\beq
\alpha_s^2 \ln{\frac{ET}{M^2}}
\longrightarrow  \alpha_s(M^2) \alpha_s(ET) \ln{\frac{ET}{M^2}} \, .
\label{rule3}
\eeq
Since $M \gg m_D$ and we already assumed \eq{approx}, 
the uncertainty when choosing $|{\tilde u}|$ as the scale in $\alpha_s$ 
is of relative order $\alpha_s(M^2) \ln{ET/M^2} \ll 1$ 
compared to the contribution $\propto \alpha_s^2 c(n_f)$ in \eq{qcd-tu}.

After we have shown how the leading logarithmic terms in the fixed coupling result 
\eq{qcd-tu} are modified when the running is taken into account (see \eq{rule2}
and \eq{rule3}), we now discuss the modification of the constant term 
$\propto \alpha_s^2 c(n_f)$. Since this term 
is the difference between the full result \eq{qcd-tu} and the leading logarithmic terms, 
it should be clear from section \ref{IIB} that it is determined by integrals over $t$ 
dominated by either $|t| \sim m_D^2$ or $|t| \sim ET$, and by integrals over $\tilde u$
dominated by $|\tilde u| \sim ET$ \footnote{The absence of
$\tilde u$-integrals dominated by $|\tilde u| \sim M^2$ (after subtraction of the logarithmic term
$\propto \ln{ET/M^2}$) is due to the fact that the $u$-channel amplitude 
squared is {\it exactly} $\propto 1/|\tilde u|$ when $|\tilde u| \sim M^2$, see the last term of
\eq{Qgampsquared}. This is in contrast with the $t$-channel amplitude squared in the region 
$|t| \sim m_D^2$, which is not exactly $\propto 1/t$ (when weighted by $t$), because of HTL complications. 
The latter lead to a remnant contribution dominated by $|t| \sim m_D^2$ after subtraction of the 
$t$-channel logarithmic term $\sim \ln{ET/m_D^2}$.}. 
There is no logarithmic spread there. Thus with running coupling, the term $\alpha_s^2 c(n_f)$ 
should be evaluated at a scale chosen {\it arbitrarily} 
\footnote{Splitting the $\alpha_s^2 c(n_f)$ term as $c_1 \alpha_s^2(m_D^2) +c_2 \alpha_s^2(ET)$
is an easy task in our calculation. However, the distinction between $\alpha_s^2(m_D^2)$
and $\alpha_s^2(ET)$ would be meaningless since we neglect contributions of order 
$\alpha_s^3 \ln{ET/m_D^2}$.} between $m_D^2$ and $ET$.

\begin{widetext}
We can now give the result for $dE/dx$ of a fast heavy quark, at leading order in the 
running coupling, 
\beq
\frac{dE}{dx}  = \frac{4 \pi T^2}{3} 
\left[ \left( 1 +\frac{n_f}{6} \right) \alpha_s(m_D^2) \alpha_s(ET) \ln\frac{E T}{m_D^2} 
+ \frac{2}{9} \alpha_s(M^2) \alpha_s(ET) \ln\frac{E T}{M^2} 
+ \alpha_s^2 c(n_f) \right] \, .
\label{qcd-running0}
\eeq
This result has been derived for $E \gg M^2/T$ and assuming \eq{approx}. The constant $c(n_f)$ is given in \eq{cofnf}. 
As discussed above, we are free to 
set the scale in $\alpha_s^2 c(n_f)$ as $\alpha_s^2 \to \alpha_s(m_D^2) \alpha_s(ET)$, in order to rewrite
\eq{qcd-running0} as
\beq
\frac{dE}{dx}  = \frac{4 \pi T^2}{3} \alpha_s(m_D^2) \alpha_s(ET)
\left[ \left( 1 +\frac{n_f}{6} \right)  \ln\frac{E T}{m_D^2} 
+ \frac{2}{9} \frac{\alpha_s(M^2)}{\alpha_s(m_D^2)} \ln\frac{E T}{M^2} 
+  c(n_f)  + \morder{\alpha_s(m_D^2) \ln{\frac{ET}{m_D^2}}} \right] \, ,
\label{qcd-running}
\eeq
where we displayed the order of neglected terms.
\end{widetext}

As a final remark, we note that when the logarithmic term $\propto \ln{E T/m_D^2}$ from $t$-channel exchange is dominant in 
\eq{qcd-running}, we obtain an interesting relation between the fast heavy quark collisional loss and 
the gluon Debye mass,
\beq
\frac{dE}{dx} \simeq \frac{m_D^2}{3} \alpha_s(ET) \ln\frac{E T}{m_D^2} \, .
\label{relation}
\eeq
We used the self-consistent equation for the QCD Debye mass derived with
running coupling \cite{APscreening},
\beq
m_D^2 = 4 \pi \l( 1 +\frac{n_f}{6} \r) \alpha_s(m_D^2) T^2 \, .
\label{selfDebye}
\eeq
Using \eq{running}, the $E\to \infty$ limit of \eq{relation} yields an upper bound for $dE/dx$, 
\beq
\frac{dE}{dx} \leq \frac{m_D^2}{12 \pi \beta_0} =  \frac{4 \pi}{33-2n_f} m_D^2 \, .
\label{bound}
\eeq
For $n_f =3$ and $m_D \simeq 0.66\,{\rm GeV}$ (at $T = 0.2\,$GeV, see below), this gives the 
bound $dE/dx \leq 1.0 \,{\rm GeV/fm}$.

%==========================================================
\section{A numerical estimate}
\label{sec:numest}

Our final result \eq{qcd-running} for the fast heavy quark collisional energy loss is predictive:
the scales at which to evaluate the different factors of $\alpha_s$ are determined,
and the order of its theoretical uncertainty is known. Consider the condition \eq{approx},
in the limit of very high
energy $E$ and very high temperature $T$, such that the logarithm
$\ln{ET/m_D^2} \sim \ln{E/T}$ is kept fixed. Then the neglected contributions in \eq{qcd-running}
are indeed small, due to $\alpha_s(m_D^2) \ll 1$ in the (very) high temperature limit.
The explicit terms in \eq{qcd-running} are then all meaningful, including that $\propto c(n_f)$.

In practice, say under RHIC conditions, the values of $m_D^2$ and $ET$ are not very different 
(on a logarithmic scale) from $\Lambda^2$. 
Therefore, the condition \eq{approx} is not satisfied in the strict sense, and the
theoretical uncertainty affecting \eq{qcd-running} is rather large.
We take $n_f=3$ and the generic values (in GeV) $\Lambda = 0.2$, $T=0.2$,
$M=1.3$ (charm quark), $E=20$. We use \eq{running} and \eq{selfDebye} and find 
$m_D \simeq 0.66\,{\rm GeV}$ (in good agreement with lattice QCD calculations, cf.\ \cite{APscreening}), $\alpha_s(m_D^2) \simeq 0.58$ and $dE/dx \simeq 0.6 \,{\rm GeV/fm}$.
The separate contributions in the bracket of Eq.~\eq{qcd-running} read $[3.31 + 0.12 + 0.49]$,
to be compared to the magnitude of neglected terms $\alpha_s(m_D^2) \ln{ET/m_D^2} \simeq 1.29$.

Thus the theoretical uncertainty is larger than the collinear logarithm from
$u$-channel exchange and than the constant $c(n_f)$ in \eq{qcd-running} -- making the approximation 
\eq{relation} reasonable under RHIC conditions. In the absence of
an explicit calculation of the neglected terms, our result $dE/dx \simeq 0.6 \,{\rm GeV/fm}$
in the above conditions might be accurate only up to a factor $\sim 2$ or so, and
should be considered at best as a sound estimate.

%=========================================================
\begin{acknowledgments}
We thank Yuri Dokshitzer for a very instructive discussion and 
fruitful suggestions. 
\end{acknowledgments}
%==========================================================

%==========================================================
\end{document}